\documentclass[lettersize,journal]{IEEEtran}
\PassOptionsToPackage{hidelinks}{hyperref}
\usepackage{amsmath,amsfonts}
\usepackage{algorithmic}
\usepackage{enumitem}
\usepackage{algorithm}
\usepackage{array}
\usepackage{textcomp}
\usepackage{url}
\usepackage{verbatim}
\usepackage{graphicx}
\usepackage{cite}
\usepackage{multirow}
\usepackage{arydshln}
\usepackage{threeparttable}
\usepackage{amssymb}   
\usepackage{float}  
\usepackage{orcidlink}

\usepackage[font=normalsize,labelfont=sf,textfont=sf]{subfig}

\usepackage[utf8]{inputenc}

\usepackage{bm}
\usepackage{booktabs}
   
\usepackage[acronym]{glossaries}
\makeglossaries
\newacronym{cnn}{CNN}{Convolutional Neural Network}
\newacronym{ml}{ML}{Machine Learning}
\newacronym{dl}{DL}{Deep Learning}
\newacronym{cu}{CU}{Central Unit}
\newacronym{uwb}{UWB}{Ultra Wide Band}
\newacronym{prf}{PRF}{Pulse Repetition Frequency}
\newacronym{ai}{AI}{Artificial Intelligence}
\newacronym{6g}{6G}{sixth-Generation}
\newacronym{cw}{CW}{Continuous-Wave}
\newacronym{pri}{PRI}{Pulse Repetition Interval}
\newacronym{cdma}{CDMA}{Code Division Multiple Access}
\newacronym{rnn}{RNN}{Recurrent Neural Network}
\newacronym{rt}{RT}{Range-Time}
\newacronym{rd}{RD}{Range-Doppler}
\newacronym{har}{HAR}{Human Activity Recognition}
\newacronym{mse}{MSE}{Mean Squared Error}
\newacronym{ae}{AE}{Autoencoder}
\newacronym{svm}{SVM}{Support Vector Machine}
\newacronym{knn}{KNN}{K Nearest Neighbor}
\newacronym{rcs}{RCS}{Radar Cross Section}
\newacronym{isac}{ISAC}{Integrated Sensing and Communication}
\newacronym{sc}{SC}{Supervised Contrastive}
\newacronym{bn}{BN}{batch normalization}
\newacronym{lopo}{LOPO}{leave-one-person-out}
\newacronym{nlp}{NLP}{natural language processing}
\newacronym{ig}{IG}{Integrated Gradients}
\newacronym{t-sne}{t-SNE}{t-distributed Stochastic Neighbor Embedding}
\newacronym{dft}{DFT}{Discrete Fourier Transform}
\newacronym{xai}{XAI}{Explainable Artificial Intelligence}
\begin{document}

\title{Transparent and Resilient Activity Recognition via Attention-Based Distributed Radar Sensing}
\author{
Mina~Shahbazifar\,\orcidlink{0000-0002-4948-5825},
Zolfa~Zeinalpour-Yazdi\,\orcidlink{0000-0002-2000-6392},
Matthias~Hollick\,\orcidlink{0000-0002-9163-5989}, 
Arash~Asadi\,\orcidlink{0000-0001-9946-4793},
and
Vahid~Jamali\,\orcidlink{0000-0003-3920-7415}
\thanks{Shahbazifar, Hollick, and Jamali's work has been co-funded by the LOEWE initiative (Hesse, Germany) within the emergenCITY center [LOEWE/1/12/519/03/05.001(0016)/72].}
\thanks{M.~Shahbazifar and Z.~Zeinalpour-Yazdi are with the Electrical Engineering Department, Yazd University, Yazd 8915818411, Iran (e-mail: minashahbazifar@stu.yazd.ac.ir; zeinalpour@yazd.ac.ir).}
\thanks{V.~Jamali and M.~Shahbazifar are with the Resilient Communication Systems Group, Technical University of Darmstadt, 64283 Darmstadt, Germany (e-mail: vahid.jamali@tu-darmstadt.de; m.shahbazifar@rcs.tu-darmstadt.de).}
\thanks{M~Hollick is with the Secure Mobile Networking Lab, Technical University of Darmstadt, 64289 Darmstadt, Germany (e-mail: mhollick@seemoo.tu-darmstadt.de).}
\thanks{A.~Asadi is with the Wireless Communication and Sensing Lab, Delft University of Technology, 2628 CD Delft, The Netherlands (e-mail: A.Asadi@tudelft.nl}
}

\maketitle

\begin{abstract}
Distributed radar sensors enable robust human activity recognition. However, scaling the number of coordinated nodes introduces challenges in feature extraction from large datasets, and transparent data fusion. We propose an end-to-end framework that operates directly on raw radar data. Each radar node employs a lightweight 2D \gls{cnn} to extract local features. A self-attention fusion block then models inter-node relationships and performs adaptive information fusion.
Local feature extraction reduces the input dimensionality by up to 480×. This significantly lowers communication overhead and latency. The attention mechanism provides inherent interpretability by quantifying the contribution of each radar node. A hybrid supervised contrastive loss further improves feature separability, especially for fine-grained and imbalanced activity classes.
Experiments on real-world distributed \gls{uwb} radar data demonstrate that the proposed method reduces model complexity by 70.8\%, while achieving higher average accuracy than baseline approaches. Overall, the framework enables transparent, efficient, and low-overhead distributed radar sensing.
\end{abstract}


\section{Introduction}
\IEEEPARstart{T}he increasing demand for intelligent ambient systems has driven significant research interest in wireless activity recognition.  Various signal representation and neural network architectures have been investigated to extract meaningful and actionable insights from wireless signals\cite{liu2019wireless,zhang2023survey}. In particular, deep neural networks have received significant attention due to their capability to automatically learn discriminative features from intricate data\cite{zhang2020device}. Several signal modalities (e.g., WiFi and mmWave) have been investigated, with \gls{uwb} radar offering superior temporal and spatial resolution due to its extremely short pulses \cite{ren2015noncontact}. 
For real-world deployment of activity recognition, especially in sensitive scenarios such as high-risk operational zones in industrial environments, continuous sensing is essential to ensure that user activities can be reliably recognized regardless of location or orientation \cite{ waqar2023direction, kruse2023continuous}. In this regard, distributed sensing has emerged as a compelling solution for resilient activity recognition\cite{zhu2022continuous}. 
However, scaling up the number of coordinated sensing nodes introduces two key 
challenges: \textit{First}, additional radars produce substantially larger data 
volumes and higher preprocessing costs, particularly when forming Doppler signatures or point clouds. \textit{Second}, achieving robust and 
interpretable fusion becomes difficult because each radar node experiences 
geometry-dependent variations in signal quality due to changes in target position, 
orientation, and propagation conditions.

\textbf{Prior work.} Current solutions attempt to address these challenges, but they do so only partially. Most distributed radar systems still depend on heavy preprocessing before learning or classification, including \gls{rd} maps \cite{guendel2021continuous, zhao2022distributed,guendel2024multipath}, Doppler  spectra \cite{zhu2022continuous,rani2020action, guendel2022distributed, waqar2023direction}, and point clouds \cite{zheng2024direction,kruse2023radar}. These \gls{dft}-based pipelines require additional operations such as side lobe suppression \cite{liu2023echoes,zhang2021widar3}, which introduce extra computational overhead. Furthermore, such representations may discard useful raw information that hinders the ability to model geometry-dependent variability \cite{liu2023echoes}.

A second body of work focuses on fusion. Existing approaches typically rely on early, intermediate, or late fusion mechanisms, each with inherent limitations. Early fusion requires transmitting raw measurements from all nodes to a central processor, resulting in significant 
communication bandwidth and making the system highly sensitive to noisy or low-quality 
nodes \cite{guendel2022distributed,dey2024radar,liu2023echoes,eckrich2024fronthaul}. 
Late fusion reduces bandwidth by combining high-level decisions, but restricts 
interaction across viewpoints and shifts computational load to individual sensors, 
limiting the ability to exploit complementary information 
\cite{guendel2021continuous,kruse2023radar,zhu2022continuous,qu2025radar}. 
Intermediate fusion offers a more balanced trade-off by allowing each radar to extract 
local features before joint integration, yet the resulting representations often lack 
interpretability because the contribution of individual nodes becomes entangled 
\cite{zhu2022continuous,waqar2023direction,zheng2024direction}.

\begin{figure}[t!]
\centering
\includegraphics[width=0.49\textwidth]{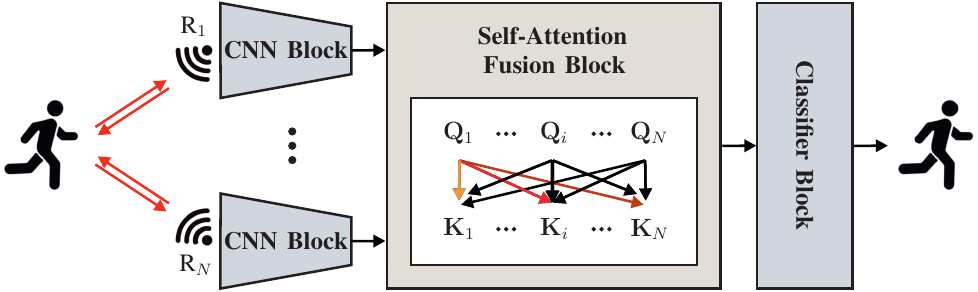}
\caption{The proposed distributed radar recognition model consists of a 2D CNN block for local feature extraction, a self-attention block for adaptive and interpretable fusion, and a final classification block.}
\label{fig:workfolw}
\end{figure}

\textbf{Our solution.}
To address these challenges, we propose an end-to-end framework that operates directly on raw radar data. The model uses lightweight 2D \gls{cnn} encoder blocks to extract local representations at each radar node. A self-attention fusion block then models inter-node relationships and adaptively integrates information from different viewpoints.
After training, the CNN encoder is deployed locally at each node. Only compact and semantically rich features are transmitted to the fusion processor. This design reduces communication overhead and inference latency.
The attention mechanism provides inherent interpretability \cite{velickovic2017graph,hu2025towards}. Attention weights explicitly quantify the contribution of each radar node to the final decision. This removes the need for post-hoc explainability methods.
In addition, a hybrid supervised contrastive loss improves feature separability. This is particularly beneficial for fine-grained and imbalanced activity classes \cite{wang2021contrastive,khosla2020supervised}.
Our main contributions are:
\begin{itemize}
\item \textbf{End-to-end raw-data framework.} 
We propose an end-to-end framework for distributed \gls{uwb} radar-based continuous activity recognition that operates directly on raw radar data, eliminating costly preprocessing like Doppler signatures and point clouds. The architecture includes a 2D CNN block for local feature extraction, a self-attention block for adaptive inter-node fusion, and a fully connected block for classification (Fig.~\ref{fig:workfolw}). Training uses a hybrid supervised contrastive loss to enhance feature separability for fine-grained, imbalanced activity classes.
\item \textbf{Interpretable attention-based multi-node fusion.} 
We introduce an attention-based fusion mechanism for distributed radar sensing. A self-attention block explicitly models inter-node relationships and produces a weight matrix that quantifies the contribution of each radar node. Ablation experiments show that the attention-derived importance scores reflect each node’s true impact on the final decision. This provides inherent interpretability that is not available in early, intermediate, or late fusion schemes. 
\item \textbf{Communication efficiency.} By deploying the trained lightweight 2D CNN encoder at each node, semantically rich compressed features are transmitted to the central processor instead of raw data which significantly reduces communication overhead and latency. We demonstrate the effectiveness of this approach compared to conventional downsampling schemes.
\end{itemize}
\section{System and Signal Model}
\textbf{System model.} A network of $N$ monostatic \gls{uwb} radar nodes is deployed around an area of interest to perform activity recognition regardless of the location and orientation of the target. The transmission of radar signals are performed in an orthogonal manner that allows simultaneous operation. Each radar transmits pulses with the same \gls{pri} and in synchronization with other nodes. The radars collect the backscattered echoes, which contain motion-induced modulation characteristics of target movements from different viewpoints. Radar nodes forward their data to a fusion processor for the recognition task.

\textbf{Signal model.} \gls{uwb} signal is typically organized as a 2D fast-time $\times$ slow-time matrix, where fast time represents range bins and slow time captures pulse-to-pulse variations, such as target motion. Let $y[n,m]$ denote the $(n,m)$-th element of this matrix, corresponding to the $n$-th fast-time sample of the $m$-th pulse. Following \cite{ren2015noncontact}, the \(m\)-th transmitted pulse, with a carrier angular frequency \(\omega_c\), can be modeled as:
\begin{equation}
x(t,mT) = p(t,mT)\,\sin(\omega_c t),
\end{equation}
where \(p(t,mT)\) is the short-duration pulse and is $T$ the \gls{pri}. The received echo at this sample, acquired with a fast-time sampling period $T_s$, is expressed as:
\begin{equation}
y[n,m] = \alpha(n,m)\,p(nT_s - t_D,mT)\,e^{j \omega_c t_D(n,m)},
\label{eq:y}
\end{equation}
where $\alpha(n,m)$ and $t_D(n,m)$ denote the attenuation factor and propagation delay, respectively. Both quantities are influenced by the target’s orientation, location, and the type of activity being performed.

\section{Proposed Framework}
 

In this section, we outline the proposed end-to-end framework. We first describe the input structure, followed by the model architecture composed of three components: \textit{(i)} a 2D CNN block for per-radar feature extraction, \textit{(ii)} a self-attention fusion block for modeling inter-node dependencies and emphasizing informative nodes, and \textit{(iii)} fully connected layers for final classification. We then present the hybrid supervised contrastive learning objective and, finally, the procedure for estimating node importance.

\subsection{Model Input Representation}
For each fixed-length slow-time window, the complex samples are converted to polar form, $\big(|y[n,m]|, \angle y[n,m]\big) $, and arranged into a three-dimensional tensor with dimensions corresponding to range bins (fast time) $\times$ window length (slow time) $\times$ 2 (magnitude/phase) which enables the model to capture both spatial and temporal variations in the scene. 
\subsection{Model Architecture}
\textbf{Feature extraction block.} The fast-time and slow-time dimensions are treated as height and width, respectively, with the polar representation as the channel dimension. Three 2D CNN layers are employed to capture spatial-temporal patterns at each radar. Each layer is followed by \gls{bn} and ReLU activation to improve optimization stability and nonlinear representational capability. A 1×1 convolutional layer is added to enhance
channel-wise feature interactions. Finally, an average pooling is employed to condense depth and spatial information into compact yet discriminative representations to enhance both computational efficiency and practical usability.

Each radar node is processed through the same block using a weight-sharing strategy\cite{zhu2022continuous}. This contributes to reduced model complexity and enhanced model scalability. The extracted features are first flattened into an array of size $d_\text{model}$ and then concatenated across the nodes to form the output $ \mathbf{S}$ with shape $N \times d_\text{model}$ compatible with the self-attention input, and forwarded to the fusion block.

\textbf{Fusion block.} Each radar observes the activity from a distinct viewpoint, depending on the target’s location and orientation, which affects the resolution and the information it acquires.
Treating the nodes as a sequence, we employ a self-attention mechanism to dynamically aggregate information across them and assign higher weights to more informative ones. At the core of this mechanism lies connecting all the nodes, which allows the model to find the hidden inter-node relation. This is achieved through three learnable linear projections that map the input features into the Query ($\mathbf{Q}$), Key ($\mathbf{K}$) and Value ($\mathbf{V}$). In this manner, the information that nodes lack can be obtained from other nodes and they complete their information from different viewpoints. By leveraging attention weights, each node focuses on the nodes that provide the most complementary information and combines their values according to the weights.
In multi-head attention with $H$ heads, rather than a single set of $\mathbf{Q}, \mathbf{K}, \mathbf{V}$, each head maintains its own set:
\begin{equation}
\mathbf{Q}_h = \mathbf{S}\mathbf{W}^Q_h, \quad 
\mathbf{K}_h = \mathbf{S}\mathbf{W}^K_h, \quad 
\mathbf{V}_h = \mathbf{S}\mathbf{W}^V_h. \quad
\end{equation}
The attention weights and outputs of each head are given by:
\begin{equation}
\quad \boldsymbol{\alpha}_h = \text{softmax}\!\left(\frac{\mathbf{Q}_h \mathbf{K}_h^{\top}}{\sqrt{d_k}}\right),  \quad 
\mathbf{S}_h = \boldsymbol{\alpha}_h \mathbf{V}_h.
\label{eq:alpha}
\end{equation}

The outputs of all heads are then concatenated and linearly projected to form the fused feature matrix, $\mathbf{S_a}$.
The attention weights across all heads are averaged to produce a matrix of shape $N \times N$ that represents the overall inter-node attention, $\bm{\alpha}$. A residual connection is added to facilitate gradient flow. The fused features are flattened into an array $\mathbf{s}_a$ to be prepared for the next block.  

\textbf{Classifier block.} The classifier consists of two fully connected layers with a dropout layer to mitigate overfitting. A softmax function produces class probabilities, and the final output $\hat{y}$ corresponds to the class with the highest predicted probability.
\subsection{Learning and Optimization}\label{sec:learning_optimization}
Inspired by \cite{wang2021contrastive}, we adopt a hybrid loss to improve robustness for fine-grained activity classes under imbalanced distributions:
\begin{equation}
\min_{\theta_c, \theta_a, \theta_d} \mathcal{L}_{\text{hybrid}} =\mathcal{L}_{\text{CE}}(y,\hat{y}) + \gamma
 \cdot \mathcal{L}_{\text{SCL}}(y,\mathbf{s}_a),
\end{equation}
where $\mathcal{L}_{\text{CE}}$ is the cross-entropy loss, and $\mathcal{L}_{\text{SC}}$ is the supervised contrastive loss over subspace $\mathbf{s}_a$, to enhance feature separability by pulling embeddings of the same class together and pushing apart those of different classes. For a batch of $B$ samples:
\begin{equation}
\label{scl}
\mathcal{L}_{\text{SC}} = \frac{1}{B} \sum_{i=1}^{B} \frac{-1}{|\mathcal{P}(i)|} \sum_{j \in \mathcal{P}(i)} \log \frac{\exp(\mathbf{z}_i \cdot \mathbf{z}_j / \tau)}{\sum_{k \neq i} \exp(\mathbf{z}_i \cdot \mathbf{z}_k / \tau)},
\end{equation}
where $\mathcal{P}(i) = \{ j \mid y_j = y_i, j \neq i \}$ is the set of positive samples sharing the same class as anchor $i$, $\mathbf{z}_i$ is the L2-normalized embedding of sample $i$, and $\tau$ is a temperature parameter controlling the distribution sharpness. The numerator $\exp(\mathbf{z}_i \cdot \mathbf{z}_j / \tau)$ encourages \textit{intra-class cohesion}, while the denominator $\sum_{k \neq i} \exp(\mathbf{z}_i \cdot \mathbf{z}_k / \tau)$ enforces \textit{inter-class separation} by penalizing similarity with negative samples \cite{khosla2020supervised}.  

\subsection{Attention Weights and Node Importance}\label{subsec:att_weight}
The weight matrix $\boldsymbol{\alpha}$ encodes pairwise node interactions \cite{velickovic2017graph,hu2025towards}. Each row, $\boldsymbol{\alpha}_{i,:}$, represents a probability distribution of the attention weights assigned by node $i$ to other nodes, while each column, $\boldsymbol{\alpha}_{:,j}$, quantifies the total attention received by node $j$, serving as a measure of its importance or attention centrality. The global importance of node $j$ is defined as:
\begin{equation}
    I(j) = \sum_{i=1}^{N} \bm{\alpha}_{i,j}.
\end{equation}
A higher $I(j)$ indicates greater influence on model decision making, a concept used in explainability analyses \cite{ying2019gnnexplainer,hu2025towards}.

\section {Experiments and Results}
In this section, we evaluate the proposed architecture. We begin by summarizing the dataset and implementation details, including network and training parameters. We then compare the model’s performance against the baseline, followed by t-SNE visualizations and an ablation study on node-attribution interpretability. Finally, we examine the model’s compression capability for real-time deployment.

\textbf{Dataset description.}
We use the public dataset from \cite{guendel2021dataset}, which includes nine human activities performed by 14 participants at varying locations and orientations, recorded using five \gls{uwb} radars with 480 fast time bins. The dataset exhibits class imbalance and activity similarities, with the following distribution: 1) walking 29.7\%, 2) stationary 14.8\%, 3) sitting down 5.1\%, 4) standing up from sitting 4.7\%, 5) bending from sitting 11.8\%, 6) bending from standing 12.9\%, 7) falling while walking 3.3\%, 8) standing up from the ground 11.6\%, and 9) falling while standing 5.3\%.

\textbf{Training and network parameters.} Different input window sizes were evaluated, and the best results were achieved with a window length of 30 which results a final input shape of $480 \times 30 \times 2$. Model parameters are detailed in in Table~\ref{tab:networks_block}.

\setlength{\dashlinedash}{2pt}  
\setlength{\dashlinegap}{2pt} 
\setlength{\tabcolsep}{6pt}     
\renewcommand{\arraystretch}{1}
\begin{table}[h!]
\centering
\begin{threeparttable}
\caption{Network Architecture}
\label{tab:networks_block}
\begin{tabular}{c c}
\hline
Block & Layer Structure \\
\hline
\multirow{7}{*}{Feature Extraction}
& Conv2D 6, kernel (7,3), padding (3,1), BN, ReLU \\
& Conv2D 8, kernel (3,3), padding 1, BN, ReLU \\
& Conv2D 6, kernel (3,3), padding 1, BN, ReLU \\
& Conv2D 6, kernel 1 \\
& AdaptiveAvgPool (5,4) \\
\hdashline
\multirow{3}{*}{Fusion} 
& Multi-head Attention 4 heads, $d_k=d_v=24$\\
& Residual Connection \\
\hdashline
\multirow{3}{*}{Classifier} 
& Dense layer 64, ReLU \\
& Dropout rate=0.3 \\
& Dense layer 9 \\
\hline
\end{tabular}
\end{threeparttable}
\end{table}
Training is performed using the Adam optimizer at a learning rate of $3\times10^{-3}$ for up to 100 epochs. Early stopping with a patience of 10 epochs is applied to prevent overfitting. The learning rate is halved if the loss doesn't improve for 10 consecutive epochs. Parameters $\gamma$ and $\tau$ are set to 1 and 0.5, respectively, based on empirical observations. A \gls{lopo} evaluation is adopted with an 4:1 train-validation split that holds one participant out as unseen test data. To mitigate class imbalance when all participants are included, data augmentation is applied using additive Gaussian noise (scale = 0.1).


\subsection{System Performance Benchmarking} 
\begin{figure}[t]
    \centering
    \footnotesize
    \begin{tabular}{@{}c@{\hspace{0.02\linewidth}}c@{}} 
        \includegraphics[width=0.48\linewidth]{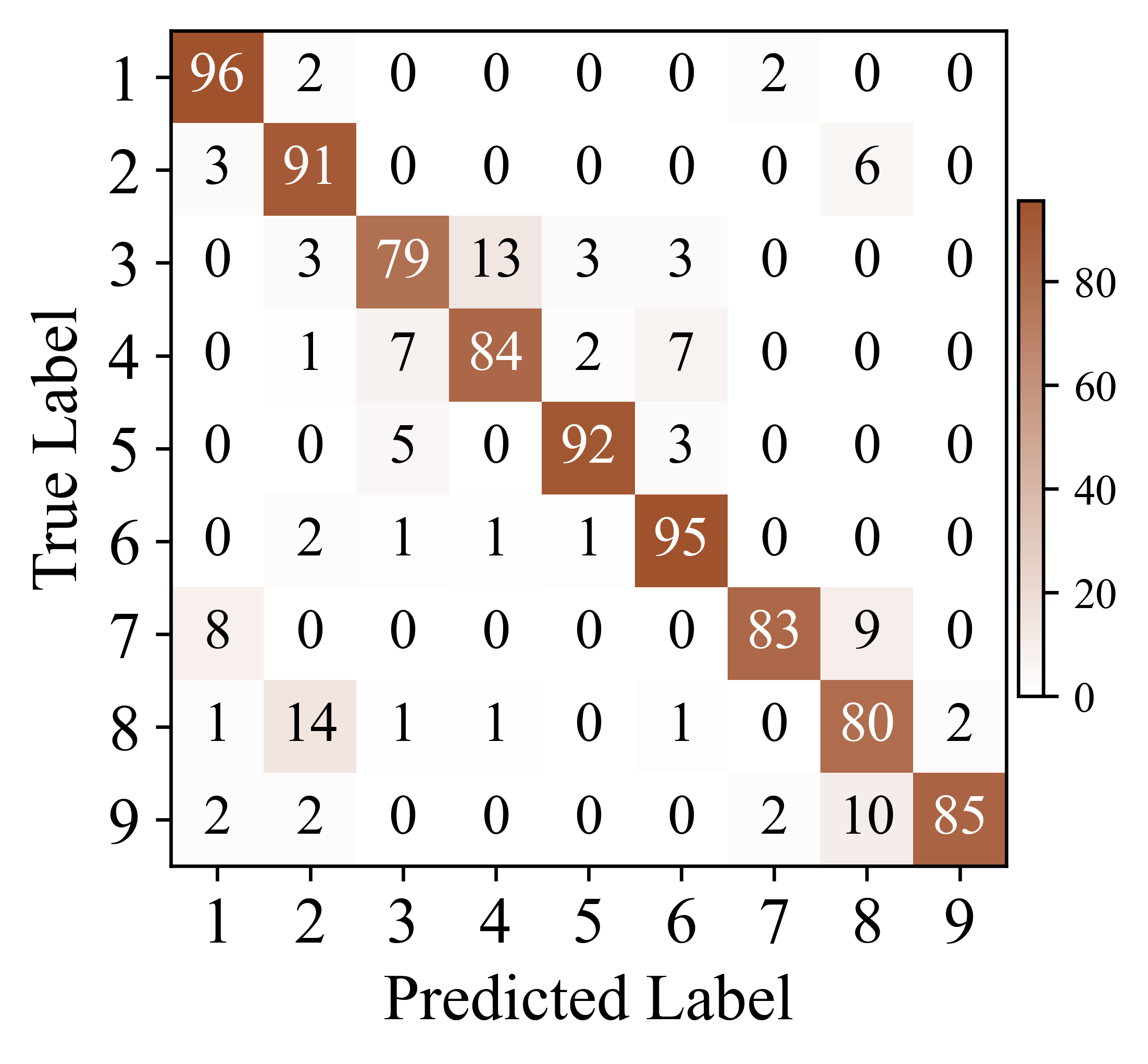} &
        \includegraphics[width=0.48\linewidth]{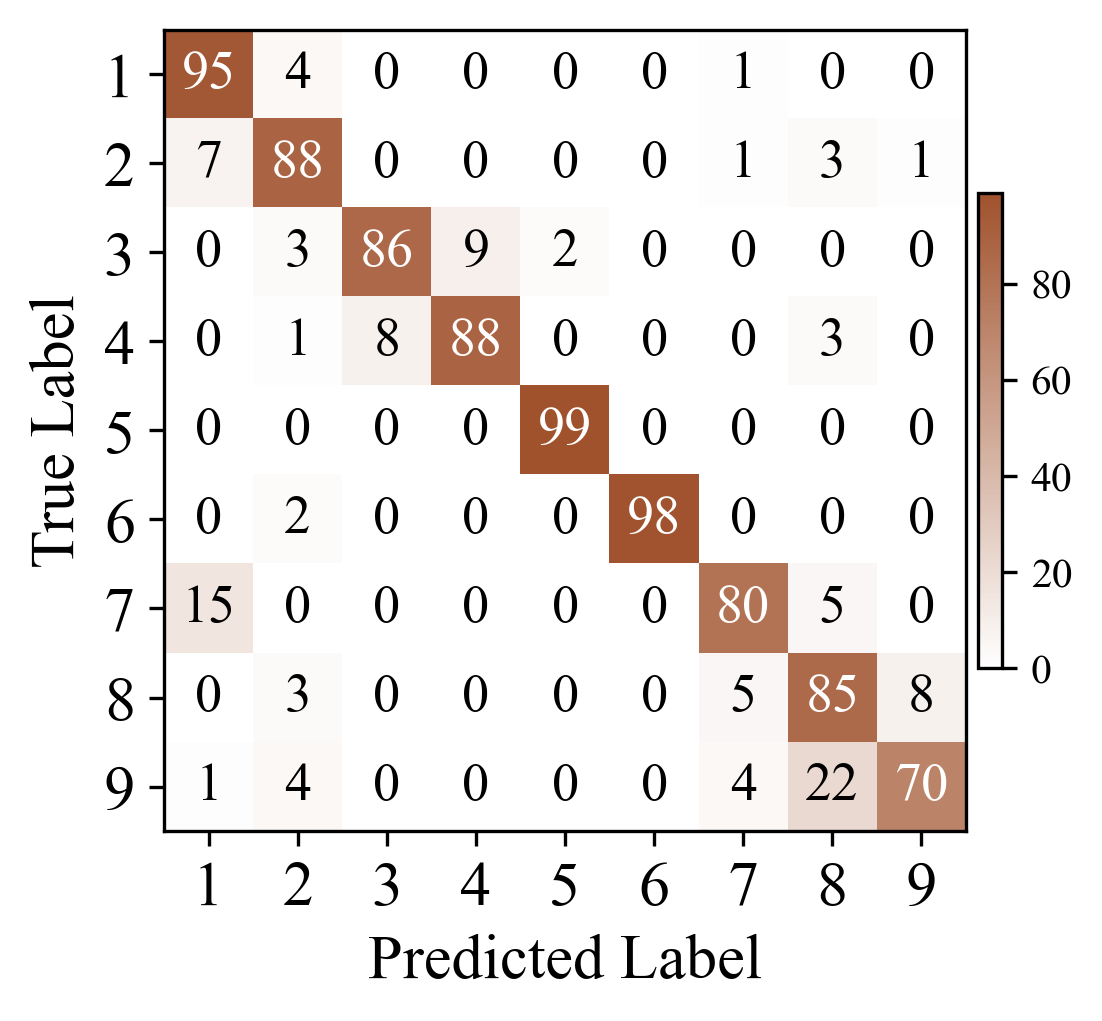} \\
        (a) &
        (b)
    \end{tabular}
    
    \caption{Confusion matrices (in percentage) for (a) the proposed model and (b) the baseline CNN-RNN\cite{zhu2022continuous}.}

    \label{fig:conf_matrices}
\end{figure}

We compare the proposed framework with the CNN–RNN model introduced in \cite{zhu2022continuous}. As shown in the confusion matrices (Fig.~\ref{fig:conf_matrices}), both models achieve higher accuracy for classes with larger sample distributions. Our model performs better in several classes (e.g., 1, 2, 7, and 9), while the baseline performs better in others (e.g., 3, 4, 5, 6, and 8). Importantly, classes with opposite movements (e.g., 3 vs. 4 and 8 vs. 9) show the highest mutual misclassifications in both models. Overall, the proposed framework shows a more favorable performance trend. As reported in Table~\ref{tab:comparison}, both the maximum and average test accuracies are improved. Notably, the proposed method reduces model complexity and eliminates costly preprocessing. Although training is more resource-intensive due to the larger dataset and hybrid optimization, this overhead is limited to offline training. In contrast, inference latency, critical for deployment, is significantly reduced: our 3.4× smaller model achieves 0.013 s per sample versus 0.049 s for the baseline on the same Intel 11th Gen Core i5-11400H CPU.
\begin{table}[t]
\centering
\caption{Performance comparison with baseline model}
\label{tab:comparison}
\resizebox{0.9\columnwidth}{!}{%
\begin{tabular}{ccccc}
\toprule
\textbf{Model} & \textbf{Max. Test} & \textbf{Ave. Test} & \textbf{Input type} & \textbf{Params} \\
\midrule
\textbf{CNN-RNN\cite{zhu2022continuous}} & 90.8\% & 85.1\% & Doppler Spectrum & 72K \\
\textbf{Proposed method} & 92.53\% & 88.54\% &  Raw data & 21K \\
\bottomrule
\end{tabular}%
}
\end{table}
\subsection{Feature Embedding Visualization}
\begin{figure}[t]
\centering
\includegraphics[width=0.48\textwidth]{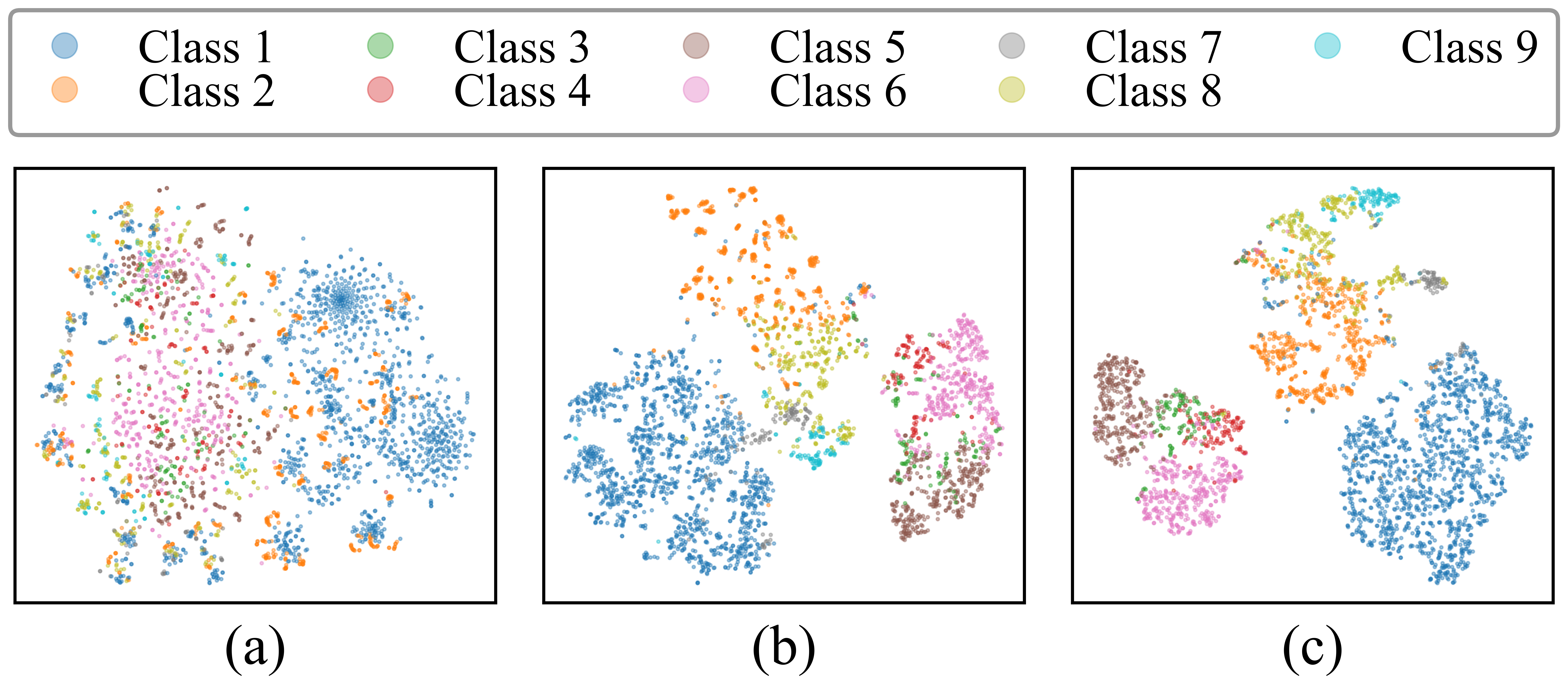}
\caption{t-SNE visualization of (a) raw input features, (b) fused representation, and (c) output of the final dense layer.}
\label{fig:t-SNE}
\end{figure}

We use \gls{t-sne} to visualize the feature embeddings. As shown in Fig.~\ref{fig:t-SNE}, plot (a) indicates substantial class overlap in the raw input data; plot (b) demonstrates clear class separation in the fused feature space that highlights effective feature discrimination. In plot (c), the classifier outputs form well-defined clusters corresponding to individual classes. Consistent with Fig.~\ref{fig:conf_matrices}(a), improved class separation in the \gls{t-sne} space (i.e., less
overlap) corresponds to higher classification accuracy in confusion matrix. For classes with overlapping movements (e.g., 7 and 9), finer segmentation could further improve both per-class and overall accuracy.
\subsection{Data Communication Overhead}
\begin{figure}[t]
\centering
\includegraphics[width=0.498\textwidth]{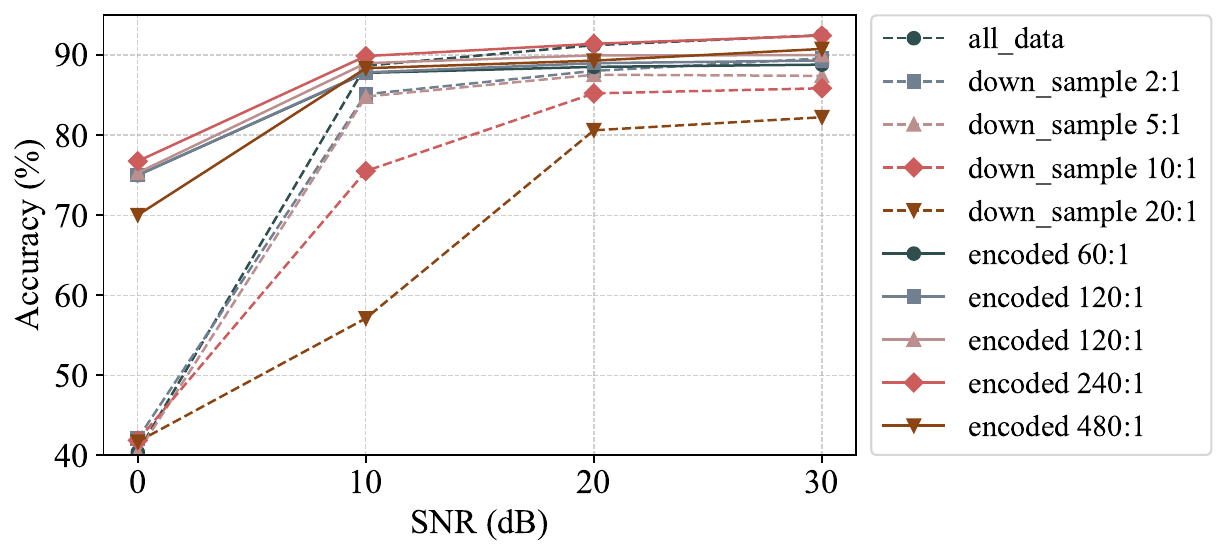}
\caption{Encoder-based feature compression versus conventional downsampling, evaluated under varying SNR conditions.}
\label{fig:compression}
\end{figure}

In real deployment, communication bottlenecks such as channel noise, latency, and limited bandwidth can substantially degrade inference performance. To address this, each radar node transmits compact features extracted by the lightweight 2D CNN block, rather than raw measurements. We benchmark this encoder-based compression scheme against conventional downsampling. Feature map sizes of (5×2), (5×4), (5×8), (10×4), and (10×8) with a fixed channel depth of 6 yield compression factors of 480×, 240×, 120×, 120×, and 60×, respectively. For comparison, fast-time downsampling ratios of 2×, 5×, 10×, and 20× are evaluated. As shown in Fig. \ref{fig:compression}, the encoder-based transmission achieves substantially higher compression and significantly better robustness to channel noise due to its extraction of noise-resilient features. Among them, the (5×4) feature map performs best. Moreover, performing part of the inference locally without additional dimensionality reduction (e.g., PCA) reduces overall latency. 
\subsection{Ablation Study and Node Importance}
\begin{figure}[t]
\centering
\includegraphics[width=0.41\textwidth]{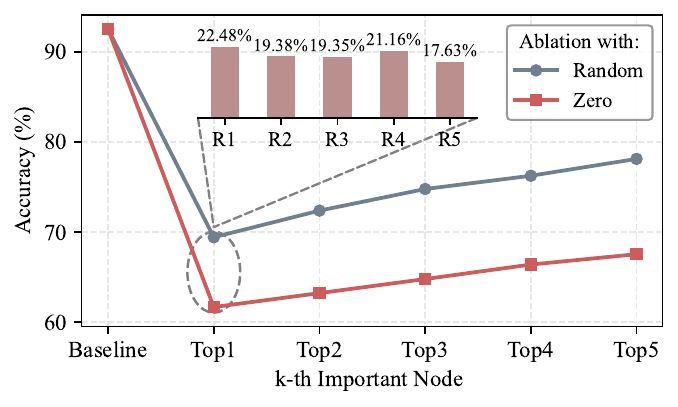}
\caption{Ablation study of node importance, with the bar plot showing the distribution of the most important nodes across the radars. }
\label{fig:ablation_single}
\end{figure}

We perform an ablation study in which portions of the input are systematically removed to evaluate the model's dependency on these inputs. Individual nodes are ablated with zeros or random data according to their computed importance from attention weights (Section~\ref{subsec:att_weight}). As shown in Fig.~\ref{fig:ablation_single}, ablating nodes with higher importance leads to larger drops in prediction accuracy. The bar plot illustrates the distribution of radars with the highest importance which indicates that the model adaptively prioritizes information from all sources. Such node-level attribution is crucial to prevent overfitting or bias during training and to ensure correct operation during the online phase (e.g., under radar failure or changes in radar orientation or placement).
\section{Conclusion}
We propose an end-to-end framework for continuous activity recognition using distributed radar that operates directly on raw data, eliminating costly DFT preprocessing. A lightweight 2D CNN extracts compact semantic features locally at each radar node, reducing communication overhead to the fusion processor. Multi-head attention–based intermediate fusion improves robustness to target position and orientation while providing interpretable node-level importance. The proposed method achieves higher accuracy with lower model complexity than baseline approaches. Ablation studies confirm the contribution of individual architectural components. The proposed feature compression strategy outperforms conventional downsampling, demonstrating its suitability for practical, low-latency distributed radar sensing.
\section*{Acknowledgment}
The authors acknowledge the compute resources provided by the high-performance computer Lichtenberg II at TU Darmstadt, funded by the German Federal Ministry of Education and Research (BMBF) and the State of Hesse.


\bibliographystyle{IEEEtran}
\bibliography{refs}
\newpage

\end{document}